\documentclass[aip, rsi, reprint, superscriptaddress]{revtex4-2}

\usepackage[utf8]{inputenc} \usepackage{amssymb,amsmath} \usepackage{graphicx}
\usepackage{braket} \usepackage{siunitx} \usepackage{xcolor} \usepackage{pgf}
\usepackage{hyperref} \usepackage{orcidlink} \usepackage{multirow}

\hypersetup{
    colorlinks=true, linkcolor=blue, citecolor=blue, filecolor=blue,
    urlcolor=blue
}

\usepackage[normalem]{ulem}

\def\Na23{^{23}\mathrm{Na}} \def\Rb87{^{87}\mathrm{Rb}}

  \def\ez{\mathbf{e}_z}

\def\muK{\mathrm{\mu K}} \def\Isat{I_{\mathrm{sat}}} \def\inch{^{\prime\prime}}
\def\mT_cm{\mathrm{mT\,cm^{-1}}} \def\T_m{\mathrm{T/m}}

\renewcommand{\arraystretch}{1.5}

\begin{document}

\title{Efficient production of sodium Bose-Einstein condensates in a hybrid
trap} \author{Y.~Geng~\orcidlink{0009-0009-1937-409X}} \thanks{These authors
contributed equally to this work.} \affiliation{Joint Quantum Institute,
University of Maryland and National Institute of Standards and Technology,
College Park, Maryland, 20742, USA}

\author{S.~Mukherjee~\orcidlink{0000-0003-0217-0743}} \thanks{These authors
contributed equally to this work.} \affiliation{Joint Quantum Institute,
University of Maryland and National Institute of Standards and Technology,
College Park, Maryland, 20742, USA}

\author{S.~Banik~\orcidlink{0000-0002-4282-3588}} \affiliation{Joint Quantum
Institute, University of Maryland and National Institute of Standards and
Technology, College Park, Maryland, 20742, USA} \affiliation{Lightmatter Inc.,
California 94040, USA}

\author{M.~Gutierrez~Galan} \affiliation{Joint Quantum Institute, University of
Maryland and National Institute of Standards and Technology, College Park,
Maryland, 20742, USA} \affiliation{IonQ, College Park, Maryland, 20742, USA}

\author{M.~Anderson~\orcidlink{0000-0001-7946-4227}} \affiliation{Joint Quantum
Institute, University of Maryland and National Institute of Standards and
Technology, College Park, Maryland, 20742, USA}

\author{H.~Sosa-Martinez~\orcidlink{0000-0002-1855-8016}} \affiliation{Joint
Quantum Institute, University of Maryland and National Institute of Standards
and Technology, College Park, Maryland, 20742, USA} \affiliation{Capital One,
McLean, Virginia 22102, USA}

\author{S.~Eckel~\orcidlink{0000-0002-8887-0320}} \affiliation{National
Institute of Standards and Technology, Gaithersburg, Maryland, 20899, USA}

\author{I.~B.~Spielman~\orcidlink{0000-0003-1421-8652}}
\email{ian.spielman@nist.gov}

\author{G.~K.~Campbell~\orcidlink{0000-0003-2596-1919}}
\email{gretchen.campbell@nist.gov} \affiliation{Joint Quantum Institute,
University of Maryland and National Institute of Standards and Technology,
College Park, Maryland, 20742, USA} \affiliation{National Institute of
Standards and Technology, Gaithersburg, Maryland, 20899, USA}

\date{\today}

\begin{abstract}

We describe an apparatus that efficiently produces $^{23}$Na Bose-Einstein
condensates (BECs) in a hybrid trap that combines a quadrupole magnetic field
with a far-detuned optical dipole trap.  Using a Bayesian optimization
framework, we systematically optimize all BEC production parameters in modest
sized batches of highly correlated parameters.  Furthermore, we introduce a
Lagrange multiplier-based technique to optimize the duration of different
evaporation stages constrained to have a fixed total duration; this enables the
progressive creation of increasingly rapid experimental sequences that still
generate high quality BECs.  Taken together, our techniques constitute a
general approach for refining and accelerating sequence-based experimental
protocols.

\end{abstract}

\maketitle

\section{Introduction}

Since their first experimental realization in 1995
\cite{anderson1995observation,davis1995bose}, atomic Bose-Einstein condensates
(BECs) have become a workhorse system for studying a diverse range quantum
many-body phenomena\cite{bloch2008many,bloch2012quantum,schafer2020tools}
including: open quantum systems \cite{ritsch2013cold}, quantum gases in low
dimensions
\cite{cazalilla2011one,langen2015ultracold,navon2021quantum,amico2022colloquium,tononi2023low},
and cosmological as well as gravitational physics
\cite{lahav2010realization,steinhauer2014observation,steinhauer2016observation,eckel2018rapidly,munoz2019observation,kolobov2021observation,banik2022accurate,viermann2022quantum,zenesini2024false}.
This breadth of applications has been accompanied by generations of
increasingly sophisticated apparatuses, each incorporating new techniques for
cooling, control, and measurement.  Quantum gas experiments nearly universally
operate in a cyclic manner, where each repetition creates a quantum gas that is
then manipulated, measured, and in the process destroyed.  Cutting-edge
experiments require very large data sets---making short cycle times highly
desirable---and demand a high degree of stability with minimal drift in
experimental conditions.

The overall cycle time is often dominated by the preparation process, which can
take tens of seconds.  Once the quantum gas is created, the subsequent physics
experiment can take anywhere from a few microseconds to many tens of seconds.
As a result, the preparation time is often the bottleneck for employing quantum
degenerate gases in both applied and fundamental applications.  An increased
data rate enables more effective exploratory studies, yields larger datasets
for reduced statistical noise, lowers Dick sampling noise~\cite{Santarelli1998}
for precision measurements, and reduces sensitivity to systematic drifts.
Thus, quantum gas production techniques that reduce production time and
increase stability are essential for further applications of this platform.  In
this paper, we begin by briefly describing our experimental apparatus for
generating $\Na23$ BECs; we then detail several strategies for optimally
creating BECs while simultaneously minimizing the production time.

Quantum degenerate gases are almost always produced in a two-stage process
consisting of laser cooling followed by evaporative cooling.  All such
experiments include a magneto-optical trap (MOT) as part of the laser cooling
process to prepare an initial sample of cold atoms.  In some cases, such as
ours, this MOT collects atoms from  a cold atomic beam (e.g., from a Zeeman
slower, a 2D-MOT, or a buffer gas source), while in others atoms are captured
directly from the low velocity tail of a dilute room-temperature vapor.  In our
experiment, the MOT temperature is fairly high (approximately $290\ {\rm \mu
K}$), so we include a sub-Doppler cooling stage that reduces the temperature to
about $\approx 40\ \mu{\rm K}$ before transferring the atoms to a conservative
potential (here, a magnetic trap) for the evaporative cooling stage.
Sub-Doppler cooling in $\Na23$ leaves the sample well above the typical
$\approx 5\ \mu{\rm K}$ BEC transition temperature, thereby requiring efficient
evaporative cooling.  Evaporative cooling selectively removes the most
energetic atoms while elastic collisions rethermalize the remainder at a
reduced temperature.  Because collision rates in these dilute gases are
relatively low, evaporative cooling typically accounts for the majority of the
total production time.  A primary focus of this paper is therefore to optimize
evaporation performance subject with a constrained total evaporation time.

\begin{figure}[tb]

\providecommand{\mathdefault}[1]{#1} \input{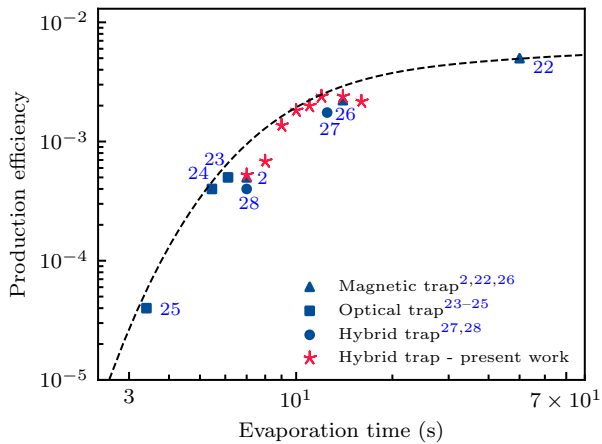}

\caption{ Comparison of different sodium BEC machines' production efficiency
and evaporation time.  Different symbols identify the underlying trapping
employed during evaporation: magnetic traps (triangles), all-optical traps
(squares), and hybrid magnetic-optical traps (circles, with our system in red).
In addition, the dashed curve serves as a guide to the eye, indicating the
nominal performance limit of existing systems in the efficiency-$t_{\rm evap}$
plane.  }

\label{fig:apparatus_compare}

\end{figure}

Given the clear delineation between initial laser cooling and the subsequent
evaporation, we quantify the evaporation stage's performance by introducing the
production efficiency $N_{\rm BEC} / N_{\rm MOT}$.  This ratio reflects how
effectively a cloud of $N_{\mathrm{MOT}}$ laser-cooled atoms is converted into
a fully condensed BEC of $N_{\mathrm{BEC}}$ atoms.
Figure~\ref{fig:apparatus_compare} plots production efficiency as a function of
the evaporation time $t_{\rm evap}$ (rather than total cycle time) for various
$^{23}\mathrm{Na}$ BEC apparatuses, thereby isolating the effect of evaporation
from any complexities in MOT loading.  Because the MOT loading time strongly
depends on variables such as the atomic source, laser power, and beam geometry,
we treat it as largely independent of evaporation optimization.  Nonetheless,
in the outlook we briefly note how coupling these two stages could provide
further performance gains.

Harmonically confining magnetic traps generated by large, out of vacuum, coils
feature very large trap volumes but with correspondingly gentle
confinement\cite{van2007large,streed2006large}.  As a result, they yield the
largest $N_{\rm BEC}$, but also require large $t_{\rm evap}$ due to their low
atom densities, and corresponding low collision rates.  These low densities
decrease 2- and 3-body loss processes, contributing to their high efficiencies.

In contrast, purely optical traps\cite{Mimoun2010, Jiang2013, Shi2021,
Liu_2021} achieve very strong confinement; while this enables rapid
evaporation, the resulting BECs are small due to limited trap volume, and
enhanced three-body losses.  Similarly in $\Rb87$, magnetic chip
traps\cite{hansel2001bose, farkas2014production} offer strong confinement in a
compact geometry, and generate small BECs with cycle times as short $1\ {\rm
s}$.

Hybrid traps, initially developed for $\Rb87$, merge the strengths of magnetic
and optical dipole traps (ODTs): first capturing atoms in a large-volume
magnetic trap, and subsequently transferring them to a tightly confining
optical trap~\cite{Lin2009,PhysRevA.93.023421,farolfi2021spin,Heo2011}.  In
Fig.~\ref{fig:apparatus_compare}, they occupy an intermediate performance
regime, achieving large $N_{\rm BEC}$ together with modest $t_{\rm evap}$.
Notably, Ref.~\onlinecite{davis1995bose} employed an optically plugged
quadrupole magnetic trap that achieved similar performance, but was rapidly
abandoned owing to alignment instability.  In these cases the same higher
density that reduces the evaporation time also accelerates 2- and 3- body loss
processes.

For all of these BEC production approaches above, BECs are produced through a
sequence of experimental stages each with a set duration, during each of which
parameters---quantifying the power and frequency of optical, microwave and RF
electromagnetic fields, the strength and direction of (near) DC magnetic
fields, to name just a few---follow preselected ramps.  Efficient BEC
production therefore requires not only a carefully chosen hardware
configuration, but also careful optimization of the stage durations and
parameter values.  After manually establishing a preliminary parameter set, we
employ a Bayesian optimization framework {\tt M-LOOP} introduced in
Ref.~\onlinecite{wigley2016fast} that is known to be performant for cold atom
experiments\cite{wigley2016fast, vuletic2022, thulium2020, Roussel2024}.  In
isolation Bayesian optimization suffers from the ``curse of dimensionality,''
leading to slow convergence (or none at all) when more than a handful of
parameters are involved.  We address this by measuring the parameters'
covariance matrix, grouping strongly correlated parameters, and then employing
Bayesian optimization to these groups sequentially.

This approach is effective in generating the globally optimal parameter set;
however, it is unable to support algebraic constraints, such as optimizing the
duration of each experimental stage subject to the constraint of fixed overall
sequence duration.  We therefore developed a Lagrange multiplier scheme to
enforce such constraints, enabling us to optimize the range of fixed-duration
evaporation sequences whose performance is shown in
Fig.~\ref{fig:apparatus_compare}.  Combining these approaches, we produced
large BECs of $4.2(1) \times 10^6$ atoms, starting with $1.8(1) \times 10^9$
atoms in the MOT, after $12\ {\rm s}$ of evaporation, and still yielded $9.2(1)
\times 10^5$ atom BECs using just 7 seconds of evaporation.

The remainder of the manuscript is organized as follows.  In
Sec.~\ref{sec:system}, we provide an overview of our experimental design at the
hardware level.  Then in Sec.~\ref{sec:stages}, we outline our BEC production
sequence.  Next, Sec.~\ref{sec:optimization} details our optimization protocol.
Finally, we conclude in Sec.~\ref{sec:conclusion}.

\section{System Description}\label{sec:system}

Here we provide a high-level description of our apparatus including
descriptions of the mechanical and optical setups.  Further details can be
found in Ph.D. thesis of S.~Banik\cite{Banik2021} and
M.~Gutierrez~Galan\cite{Gutierrez-Galan2021}.

\subsection{Vacuum System}

\begin{figure}[t]

\includegraphics{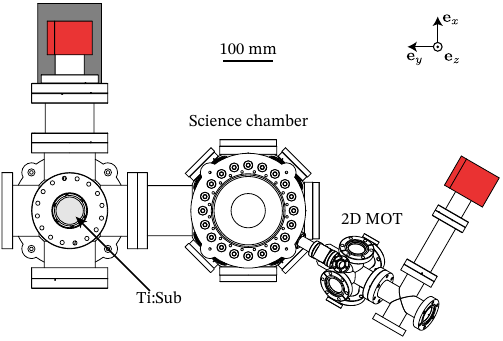}

\caption{Top view of vacuum system.  Red/gray colored regions denote vacuum
pumps with a standard ionization pump in dark gray, combination
ionization/getter pumps in red, and a Ti:Sub pump in light gray.  }

\label{fig:chamber_top_view}

\end{figure}

The vacuum system (Fig.~\ref{fig:chamber_top_view}) comprises a ``high''
pressure 2D MOT source, connected via a low conductance link to a low-pressure
UHV science chamber (giving a vacuum-limited lifetime of magnetically trapped
atoms $\approx 25~\rm {s}$).

\begin{figure}[t]

\includegraphics{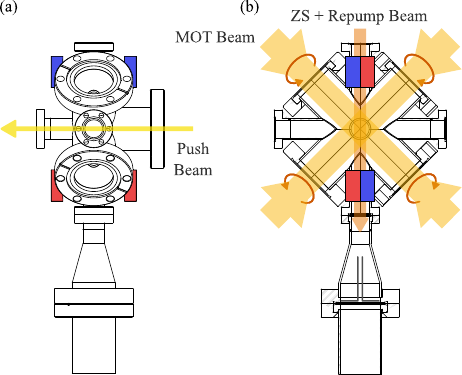}

\caption{ 2D MOT.  (a) Front view. Laser beams are colored in yellow.  The
north and south poles of magnets are colored in red and blue, respectively.
(b) Side-section view.  The polarization is marked on two MOT beams.  Zeeman
slower beam is polarized perpendicular direction of the drawing.  }

\label{fig:2d_mot}

\end{figure}

{\it 2D MOT chamber} -- Our 2D MOT (Fig.~\ref{fig:2d_mot}), inspired by
Ref.~\onlinecite{Lamporesi2013}, is built around a ten-way ConFlat cross with
four $2.75\inch$ and four $1.33\inch$ flanges in the cross's plane, plus
individual $2.75\inch$ and $1.33\inch$ flanges on the axis normal to that plane
($1\inch = 2.54\ {\rm cm}$).  The $2.75\inch$ out-of-plane flange connects to
the pumping section, and the  $1.33\inch$ flange connects to science chamber.
Finally, a gate valve and a $3~\mathrm{mm}$-diameter differential pumping tube
separate the atomic source region from the science chamber.  A
resistively-heated sodium oven is mounted on the bottom in-plane $1.33\inch$
flange.

The 2D MOT's quadrupole field is produced by four stacks of nine $25\ {\rm mm}
\times 10\ {\rm mm} \times 3\ {\rm mm}$ neodymium bar magnets [Eclipse N750-RB
\footnote{Certain equipment, instruments, software, or materials are identified
in this paper in order to specify the experimental procedure adequately.  Such
identification is not intended to imply recommendation or endorsement of any
product or service by NIST, nor is it intended to imply that the materials or
equipment identified are necessarily the best available for the purpose.}, with
magnetization $(8.8 \pm 0.1) \times 10^5$~A~m$^{-1}$] mounted on the
cross\cite{Lamporesi2013, Tiecke2009}.  The stacks are divided into front and
back sets spaced by $70\ {\rm mm}$, and each set contains two stacks that are
vertically separated by $98~\mathrm{mm}$.  These result in a 2D quadrupole
field with a $0.42\ {\rm T}/{\rm m}$ gradient; this gradient changes by only
$\approx 10\ \%$ over the longitudinal extent of the 2D MOT.

{\it Science chamber} -- The science chamber (Kimball Physics
MCF800-SphSq-G2F1E3C4A16) is an $8\inch$ stainless-steel vessel with recessed
windows on the top and bottom.  Four $4.5\inch$ and four $2.75\inch$ flanges
reside on the horizontal plane and eight pairs of $1.33\inch$ flanges are
placed $21^\circ$ above and below the horizontal plane.  The recessed windows'
inner surfaces are setback by $20\ {\rm mm}$ from the chamber's center.

Magnetic field gradients for the 3D MOT and magnetic trap are generated by a
pair of coils in an anti-Helmholtz configuration.  Each coil is wound from
Kapton insulated $3/16\inch$ ($0.48\ {\rm cm}$) hollow square copper tube and
is mounted just outside a recessed window.  At the peak current of
$200~\mathrm{A}$, these coils provide a $2.2~\T_m$ gradient along $\ez$ and are
water-cooled with a flow rate of $0.6~\mathrm{L\,min^{-1}}$, supplied by a
booster pump operating at $1.25~{\rm MPa}$ ($180~\mathrm{psi}$).  Finally,
three additional pairs of coils in a Helmholtz configuration provide bias
magnetic fields along each Cartesian axis.

The 2D MOT chamber is actively pumped by an hybrid ion-getter pump (SAES
D-100), and the science chamber is pumped by a combination of an ion pump
(Gamma Vacuum 45S) and an hybrid ion-getter pump(SAES D-500).  A titanium
sublimation pump is attached to the science chamber but was only activated
during the initial pump-down phase.

\subsection{Optical sources} \label{sec:lasers}

\begin{table}[tb!]

    \centering \renewcommand{\arraystretch}{1.2} \setlength{\tabcolsep}{6pt}
    \resizebox{\columnwidth}{!}{

    \begin{tabular}{lcc}
	\hline\hline

	\multicolumn{3}{c}{\textbf{MOT loading}} \\

	\hline

	\textbf{Beam} & \textbf{Frequency} & \textbf{Intensity} \\

	\hline

	2D MOT cooling    & $f_{2\to3} - 1.2\Gamma$   & $1.9 \Isat$ \\

	2D MOT repump  & $f_{1\to2} - 1.5\Gamma$ & $30.5 \Isat$ \\

	Zeeman slower       & $f_{2\to3} - 18.6\Gamma$  & $23.1 \Isat$ \\

	Push beam      & $f_{2\to3} + 0.3\Gamma$   & $0.05 \Isat$ \\

	3D MOT cooling    & $f_{2\to3} - 1.4\Gamma$   & $1.7 \Isat$ \\

	3D MOT repump  & $f_{1\to2} - 1.1\Gamma$ & $1.9 \Isat$ \\

	\hline

	\multicolumn{3}{c}{\textbf{sub-Doppler cooling}} \\

	\hline

	\textbf{Beam}    & \textbf{Frequency} & \textbf{Intensity} \\

	\hline

	3D MOT cooling         & $f_{2\to3} - 3.4\Gamma$      & $0.3 \Isat$ \\

	Bright repump       & $f_{1\to2} - 1.1\Gamma$  & $0.1 \Isat$ \\

	\hline

	\multicolumn{3}{c}{\textbf{Optical pumping}} \\

	\hline

	\textbf{Beam}         & \textbf{Frequency} & \textbf{Intensity} \\

	\hline

	3D MOT cooling    & $f_{2\to3} - 0.9\Gamma$ & $0.3 \Isat$ \\

	Optical pumping  & $f_{1\to1} + 0.26\Gamma$             & $0.007 \Isat$
	\\

	\hline\hline

    \end{tabular} }

    \caption{Set of laser parameters during each laser cooling stage.
    Frequencies are reported with respect to that of the desired transition
    (i.e., $f_{2\to 3}$ denotes the frequency of the $\ket{F =
    2}\!\to\!\ket{F^\prime = 3}$ transition). Intensities are reported with
    respect to the saturation intensity of the $\ket{F = 2}\!\to\!\ket{F^\prime
    = 3}$ transition.  The red detuning of the 3D MOT- and bright-repumps is a
    legacy of initial manual optimizations that found these values to be
    insensitive at the scale of a few linewidths.  }

    \label{tab:laserCoolingStages}

\end{table}

All the laser light used for laser cooling, imaging, and repumping is derived
from two lasers resonant with distinct transitions within $\Na23$'s $589\ {\rm
nm}$ ${\rm D}_{2}$ line.  The first, a Toptica DL-RFA-SHG pro ($1.1\ {\rm W}$
output power), addresses the $\ket{F\!=\!2} \to \ket{F'\!=\!3}$ cooling
transition.  The second, a Toptica TA-SHG pro ($0.9\ {\rm W}$ output power),
couples to the $\ket{F\!=\!1} \to \ket{F'\!=\!2}$ repumping transition.  The
cooling laser is locked to a Doppler-free saturated absorption feature, while
the repump laser is beat note-locked to the cooling laser with a $1.8\ {\rm
GHz}$ frequency shift.  All laser light is delivered to the vacuum system via
single-mode optical fiber.  The nominal laser parameters are summarized in
Table~\ref{tab:laserCoolingStages}.

{\it 2D MOT lasers} -- Our 2D MOT utilizes in-plane ``cooling'' beams, a repump
beam, a Zeeman slower beam and a transverse push beam.  The two retro-reflected
cooling beams ($\approx 6\ {\rm mm}$ waist) enter and exit the 2D MOT chamber
via the four in-plane $2.75\inch$ viewports.  The repump and the Zeeman slower
beams are overlapped, enter the chamber through the top $1.33\inch$ viewport,
and pass through the 2D MOT before entering the Na oven.  Lastly, the push beam
has a $1\ {\rm mm}$ waist and enters the chamber by the longitudinal axis
$2.75\inch$ viewport (i.e. oriented along the longitudinal axis of the 2D MOT
directed into the science chamber).

{\it MOT lasers} -- A 2-in, 6-out fiber beam splitter (Evanescent Optics) then
distributes light injected into a single fiber to the six outputs that source
the 3D MOT / sub-Doppler cooling beams with waist $\approx 12\ {\rm mm}$).
Repumping during dark SPOT MOT operation is provided by a single beam (with
waist $\approx 8\ {\rm mm}$) with a $9\ {\rm mm}$ diameter dark spot imaged
onto the MOT center.  In addition, a single ``bright'' (i.e., no dark spot)
repump illuminates the atoms during sub-Doppler cooling.

{\it Optical pumping laser} -- The optical pumping beam is derived from the
same laser system as the MOT repump light, frequency-shifted by an
acousto-optical modulator to address the $\ket{F = 1}\!\to\!\ket{F^\prime = 2}$
transition. It is sent along the quantization axis set by a uniform bias
magnetic field $\approx 0.1\ {\rm mT}$ with a waist of approximately $4\ {\rm
mm}$ and power 40 $\mu{\rm W}$.

{\it ODT laser -- } The $1064\ {\rm nm}$ ODT beam is produced by Ytterbium
fiber laser (IPG YLR20-1064-LP-SF).  After pre- and post-conditioning along
with coupling through a polarization-maintaining photonic crystal fiber (NKT
LMA-PM-10) with $\approx 70\ \%$ efficiency, $6.3\,\mathrm{W}$ of power is
delivered to the vacuum system.  The beam then enters through the $4.5\inch$
viewport on the $y$-axis is focused to a $1/e^2$ waist of $22\ \mu{\rm m}$,
forming an ODT with a maximum depth of $200\ \mu{\rm K}$.

\subsection{UV desorption source}

We employ light-induced atomic desorption (LIAD) to inhibit Na buildup on the
top $2.75\inch$ 2D MOT windows\cite{gozzini1993light,liad_sodium_pra}.  This
light is sourced from a pair of LED lamps (Thorlabs M365LP1) and is merged into
the MOT beams via dichroic beam splitters yielding $30\ \rm{mW}$ at each
viewport.

\section{System operation}\label{sec:stages}

In brief, our BEC production begins with a 2D-MOT that generates a cold atomic
beam from a hot effusive $\Na23$ source.  The cold beam then travels into the
science chamber where $\Na23$ atoms are captured into a dark spontaneous-force
optical trap (SPOT~\cite{Ketterle1993}, although duplicative we will conform to
the standard language of ``dark SPOT MOT'').  Afterward they are sub-Doppler
cooled~\cite{lett1988observation}, and finally optically pumped into the
magnetically trappable low field seeking state $\ket{F=1, m_F = -1}$.

We abruptly turn on a quadrupole magnetic trap, where the atoms undergo
rf-evaporation.  Once the cloud becomes sufficiently cold and dense, it is
transferred into a horizontally aligned single-beam ODT~\cite{Lin2009}.  Forced
evaporation by lowering the ODT power brings the sample to quantum degeneracy.

\subsection{2D MOT}

The Na oven, operating at $\approx 240\ {\rm ^\circ C}$, directs an up-going
atomic beam into the 2D MOT.  Owing to Na's low vapor pressure this can rapidly
coat the top $1.33\inch$ viewport, which we prevent by continuously heating the
top viewport to $130\ ^\circ{\rm C}$.  Over longer timescales Na also
accumulates on the in-plane $2.75\inch$ 2D MOT cooling viewports.  We prevent
this accumulation via nightly exposure to the UV LIAD light.

The decreasing fringe field from the 2D MOT permanent magnets resembles the
field profile of a traditional Zeeman slower as it rises from $0\ {\rm mT}$ to
a maximum of $20\ {\rm mT}$ over a distance of about $50\ {\rm mm}$.  Motivated
by this observation~\cite{Lamporesi2013}, our cooling process begins by
Zeeman-slowing the Na atomic beam using a red-detuned down-going slower beam.
The slowed atomic beam is then captured and cooled into the 2D MOT before being
directed towards the 3D MOT by the blue-detuned push beam.  This configuration
relies on a single repump beam---co-propagating with the slower beam and
detuned by $9\ {\rm MHz}$ from the $\ket{F=1}\!\to\!\ket{F'=2}$
transition---for both the slowing and 2D cooling stages.  We operate this beam
at fairly high power to provide effective repumping for both, given the poor
transmission of the viewport.

\subsection{Dark SPOT MOT}

After entering the science chamber, the cold atomic beam from the 2D MOT is
captured by the dark SPOT MOT.  During this stage, the quadrupole magnetic
field has a $z$-axis gradient of $0.06~\T_m$.  The three pairs of
counterpropagating MOT beams are red-detuned from the $\ket{F=2}\! \to\!
\ket{F'=3}$ cycling transition and the dark-SPOT repump beam are red-detuned
from the $\ket{F=1}\!\to\!\ket{F'=2}$ transition.  In the dark volume at the
interaction of these beams atoms accumulate in the ground state $\ket{F=1}$
manifold reducing the radiation pressure and increasing the maximum atomic
density~\cite{Ketterle1993}.

All together we observe an initial atom loading rate of $0.8\times 10^9\ {\rm
s}^{-1}$ and the MOT number saturates at $2 \times 10^9$ after about $10\ {\rm
s}$.  In practice we load the MOT for $7\ {\rm s}$ yielding $1.8 \times 10^9$
atoms at $\approx 290\ \muK$.

\subsection{Sub-Doppler cooling}

After MOT loading, we perform sub-Doppler cooling by: removing the magnetic
field gradient; further red-detuning the MOT beams and reducing their
intensity; and use the bright repump beam for uniform repumping.  This is
commonly known as Polarization gradient cooling (PGC) in other literatures.
During this stage, stray magnetic fields are canceled using the bias coils.
The atoms cool to $\approx 40\ \muK$ in just $2\ {\rm ms}$ and are fully
depumped into the $\ket{F=1}$ manifold.

\subsection{Hybrid trap loading}

Prior to magnetic trapping, the sub-Doppler cooled cloud nominally uniformly
populates all three magnetic sublevels of the $\ket{F=1}$ manifold; of these
states, only the $\ket{F=1,m_F=-1}$ is a magnetically trappable weak-field
seeking state.  We therefore optically pump into $\ket{F=1,m_F=-1}$ in $1.2\
{\rm ms}$ using a low intensity (0.02 $\mathrm{mW/cm^2}$) circularly polarized
optical pumping beam driving the $\ket{F=1}\!\to\!\ket{F'=1}$, aligned to an
$\approx 0.1\ {\rm mT}$ bias field  along with depumping from the MOT
beams\cite{van2007large}.

After optical pumping, we suddenly energize the quadrupole trap to a gradient
of $0.73~\T_m$, capturing 60\% of the atoms.  The trapped ensemble is then
compressed by increasing the gradient to $1.9~\T_m$ in $240~{\rm ms}$,
adiabatically increasing the temperature to  $\approx 230~\muK$, at a nominally
fixed phase space density of $10^{-5}$ (Fig.~\ref{fig:evap_seq}d) The ODT is
then turned on to its peak power of $\approx 6\ {\rm W}$; at this time it
contributes only a small perturbation (trap depth $\approx 200~\muK$) to the
magnetic trap.

\subsection{Evaporative cooling} \label{sec:hybrid_trap}

\begin{figure}[t]

\includegraphics{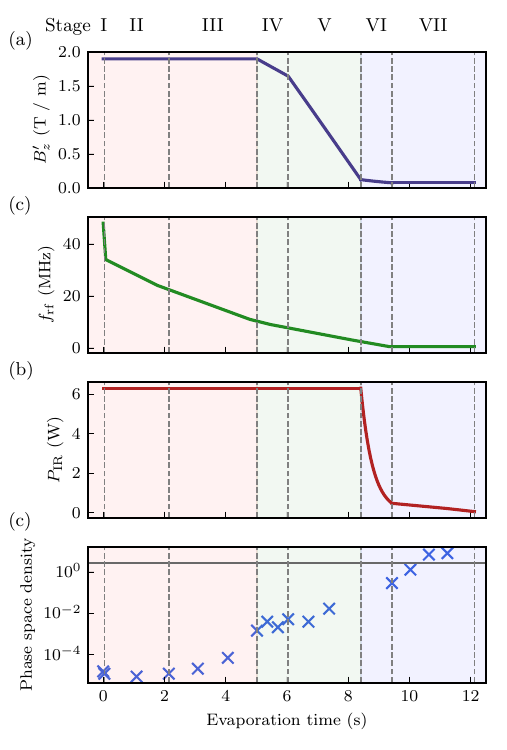}

\caption{
    The evaporation sequence.  $z$-axis quadrupole field gradient $B'_z$, rf
    frequency $f_{\rm rf}$, and ODT beam power $P_{\rm IR}$ as a function of
    time into the evaporation sequence are shown in (a), (b), and (c).  The
    sequence is divided into ramps, whose boundaries are indicated by the gray
    dash lines.  The red, green, and blue shaded regions represent the rf
    evaporation stage, the decompression/transfer stage, and the hybrid trap
    evaporation stage, respectively.  (d) shows the phase space density
    calculated from the experimental data.  The gray solids line marks the BEC
    transition's critical phase space density of $\approx 2.6$.
}

\label{fig:evap_seq}

\end{figure}

After laser cooling, we use evaporative cooling to bring the atoms to quantum
degeneracy.  Figure~\ref{fig:evap_seq} shows our optimized evaporation sequence
(see Sect.~\ref{sec:optimization}) that leads to our largest BECs.

First, we perform rf-evaporation in a purely magnetic quadrupole trap, which
uses an oscillating magnetic field.  If the atoms' Zeeman shift equals the
frequency of the rf magnetic field, their internal state is changed from
$\ket{F=1,m_F=-1}$ to $\ket{F=1,m_F=0,+1}$ and they are ejected from the trap.
At the beginning of the rf-evaporation stage, our initial frequency of 48~MHz
is quickly swept from 48~MHz to 34~MHz in 0.05~s, then slowly reduced to $11\
{\rm MHz}$ in $4.8~{\rm s}$, resulting in a cloud of approximately $50 \times
10^6$ atoms with temperature $\approx 80\ \muK$.

Near the end of the rf evaporation process we reduce the magnetic confinement
(green region in Fig.~\ref{fig:evap_seq}), allowing the ODT (placed just below
the quadruple center) to provide vertical support against gravity.  Because the
ODT contributes a dimple-like potential to the horizontal magnetic confinement,
even adiabatic transfer leads to the increase in phase-space
density~\cite{StamperKurn1998} seen in Fig.~\ref{fig:evap_seq}.  During the
hybrid trap evaporation stage (blue region in Fig.~\ref{fig:evap_seq}), the ODT
depth is ramped down from $\approx 200~\muK$ to $17~\muK$ following an
exponential ramp and a linear ramp with combined duration of $4.2\ {\rm s}$.
About half-way through this stage, the phase-space density of the atoms crosses
the critical phase density of $2.6$, at which point a BEC first
forms\cite{Pethick2001}.  The sequence concludes with an almost-pure BEC with
$4.2(1) \times 10^6$ atoms confined in a hybrid trap with frequencies:
$\omega_{\rho} = 2\pi\times 320\ {\rm Hz}$ (radial confinement from ODT) and
$\omega_{x} = 2\pi\times 22\ {\rm Hz}$ (longitudinal confinement from combined
magnetic and optical potentials).

\section{Optimization}\label{sec:optimization}

As seen in Sec.~\ref{sec:stages}, each separate experimental stage requires a
range of control and timing parameters, that are often correlated between
stages.  All of these interdependent parameters must be optimized for peak
performance.  Optimization can be abstracted as minimizing one or more ``cost
functions'' that characterize the system's performance for a particular a set
of control parameters.

A good selection of cost function requires it to be a sensitively reflects the
system performance, while also practically easy to acquire and robust to
experiment noise.  For us, we use the BEC fraction shortly after crossing the
BEC phase transition (a point in the final ramp) as the cost function, as this
value serves a sensitive agent of the phase space density.  It can be reliably
acquired by a single-shot measurement after a small time-of-flight, since the
cloud is already cold but also has a reasonably small OD, in contrast to a pure
BEC whose OD is difficult to determine.  We note that the start and end point
of the final ramp have to remain fixed for the measured BEC fraction to be
comparable.  To optimize the final ramp, we append an ``anti-evaporation" ramp
after the final ramp to adiabatically bring the system back to a fixed state,
and use the BEC fraction there as the cost function.

For experiments like ours, brute force optimization quickly becomes intractable
in the face of a large number of strongly correlated control parameters, an
associated convoluted cost function landscape, measurement noise, and of course
hardware-level technical limitations.  It is virtually impossible to construct
a top-to-bottom first-principle theoretical description of a cold-atom
apparatus; as such this optimization is largely empirical but guided by theory
and modeling.

Traditional optimization methods operate by sampling from and navigating the
cost landscape in an attempt to locate the (global) minimum.  Many contemporary
approaches now frame this as a machine learning (ML) task, with the objective
transformed into ``learning'' a high quality model of the underlying physical
reality, from which extrema can be accurately identified.  Gaussian processes
(GP), a type of Bayesian optimization, is one such approach; while this has
already yielded promising results in cold-atom
experiments~\cite{wigley2016fast,vuletic2022,Roussel2024}, many challenges
remain.  For instance, these optimization methods suffer from the curse of
dimensionality and scale poorly with increasing parameter-count in two ways:
(1) Even in principle, the required number of samples rapidly increases
(thereby increasing the requisite computational resources for Bayesian
inference); and (2) in conjunction with experimental noise, large-parameter
count GP models can often fail to converge.  Furthermore, like most ML
approaches, the performance of these methods depends strongly on the choice of
hyperparameters~\cite{sivaprasad20a}, and the mechanism (if any) of in-loop
hyperparameter tuning as the cost function landscape is explored.

We developed a multi-stage optimization procedure to circumvent the curse of
dimensionality.  This procedure begins by manual scanning each parameter to
identify an initial value and bounds for the following optimization.  Next,
these parameters are grouped into (potentially) intuitively related sets---{\it
e.g.,} ``laser cooling'' or ``ODT evaporation''---of about 10 parameters and
optimized with GP.  We then use a stochastic sampling technique to compute the
parameter correlation matrix, allowing us to quantitatively collect parameters
into sets with strong intra-set correlations and weak inter-set correlations,
which are then separately optimized.

Introducing user-defined algebraic constraints to this optimization task is a
further challenge.  This is commonly implemented by adding a penalty term to
the cost function~\cite{barker2020applying,vuletic2022}. While effective, this
technique requires careful tuning of the relative strength of the penalty with
respect to the remainder of the cost function: an excessive penalty can cause
the optimizer to diverge, while too little will be ignored.  Instead, we
constrained the optimization task by including a Lagrange multiplier in the
cost function.  When constraining the total evaporation time, such a method
enables the production of high-quality BECs with progressively shorter
experimental durations.

In the remainder of this section, we describe separate strategies to
effectively optimize large parameter sets (Sec.~\ref{sec:parameter_grouping})
and incorporate algebraic constraints
(Sec.~\ref{sec:constrained_optimization}).

\subsection{Parameter grouping} \label{sec:parameter_grouping}
\begin{figure}[t]

\includegraphics{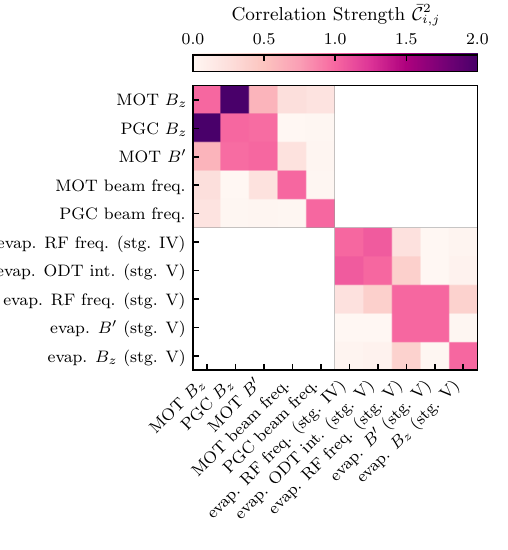}

\caption{
    Control parameter correlations.  These parameters include those used for
    laser cooling, and transfer into the hybrid trap (stage IV and V in
    Fig.~\ref{fig:evap_seq}).  The strength of correlations is represented by
    the color of the off-diagonal terms.  Each row and column is normalized by
    the value of the diagonal term (see text).
}

\label{fig:par_corr_gp}

\end{figure}

Our focus is on ML optimization tasks where experimental data acquisition rate
is low (limiting the number of samples) and computational requirements scale
poorly with the dimension of the parameter space.  In general, this situation
is hopeless and giving up is likely the most rational response.
Experimentalists are not known for such sensible behavior, and besides, in
laboratory settings even a highly non-trivial cost function landscape may
contain some hidden structure.  In our case, this structure is encoded in
correlations between control parameters: parameters that are highly correlated
in the neighborhood of one critical point (local minima, maxima, or saddle
point) are likely to be correlated near another.  This converts a hopeless
situation into a hopeful three-step process of: identifying at least one
critical point, obtaining the parameter's correlation matrix at that location,
and optimizing parameters in sets with strong correlations.

During the setup stage of any real experiment, every control parameter is
likely to have been studied at least once; for each parameter, this provides
working information regarding its nominal optimal operating point, and overall
range of utility.  Taken together, this set of nominal optimal operating points
identifies a critical point in parameter space with zero gradient of the
fitness function along all directions, therefore suitable to explore the
correlation matrix.  In practice, we employed several rounds of human-guided
manual tuning and GP optimization to identify the initial operating point,
denoted by the vector ${\bf x}^{(0)} \equiv \{\Delta x^{(0)}_0, \Delta
x^{(0)}_1, \cdots \}$ in parameter space.

We then randomly sample the cost function at points ${\bf x}^{(n)}$ in the
neighborhood of the initial operating point, i.e, small $\Delta x^{(n)}_i
\equiv x_i - x^{(0)}_i$,  and fit a second-order polynomial \begin{align}
    C(\bf{\Delta x}) &\approx \mathcal{A} + \sum_i \mathcal{B}_i \Delta x_i +
    \sum_{i,j} \Delta x_i \mathcal{C}_{i,j} \Delta x_j
\end{align} to the resulting data.  Here $\mathcal{A}$ measures the nominal
value of the cost function, $\mathcal{B}_i$ is the local gradient [ideally
zero, at $x^{(0)}_i$], and $\mathcal{C}_{i,j}$ is the desired covariance
matrix.  The diagonal elements of $\mathcal{C}_{i,j}$ quantify the individual
parameter sensitivities, and the off-diagonal elements describe
cross-correlations.  Because the parameter sensitivities are completely
arbitrary (given by whatever units happen to be used in the experimental
control system), we quantify parameter correlations in terms of squared
normalized correlation matrix $\bar{\mathcal{C}}^2_{i,j} \equiv
\mathcal{C}^2_{i,j} / (\mathcal{C}_{i,i}\mathcal{C}_{j,j})$ whose diagonal
elements have been scaled to unity.

Figure~\ref{fig:par_corr_gp} illustrates $\bar{\mathcal{C}}^2_{i,j}$ for two
sets of parameters taken from two very different parts of the experimental
sequence: the first set includes parameters from the laser cooling stages,
while the other contains parameters from the hybrid trap transfer stage (stage
IV and V in Fig.~\ref{fig:evap_seq}).  As perhaps intuitively expected, only
parameters within these sets are correlated.  However, even within these sets,
specific parameters exhibit differing degrees of correlation.

{\it Laser cooling parameters}---the bias and gradient magnetic fields during
the MOT loading and sub-Doppler cooling stages are strongly correlated; this
arises from the slow response time of the coils.  By contrast, the laser
detunings during the MOT loading and sub-Doppler cooling stages control very
different processes and are weakly correlated.

{\it Hybrid trap transfer parameters}---The magnetic trap gradient and rf
frequencies are highly correlated as their interplay controls the evaporation
efficiency.  Likewise, the rf frequency and the ODT intensity are strongly
correlated because the rf frequency sets the temperature of the cloud which the
ODT must then be deep enough to capture.

In the end, experimentally identified correlations such as these are used to
partition the whole parameter set into strongly correlated subsets that are
optimized together.

\subsection{Bayesian optimization}

We employ a GP based Bayesian optimizer which is particularly effective in
settings where sampling the cost function is expensive, the number of sampled
points is restricted, and experimental noise is present.

At every optimization, a GP optimizer encodes a probability distribution over
the set of all potential cost functions $\left\{f({\bf x})\right\}$, where the
probability density of a specific cost function is given by a (functional)
normal distribution.  The task of the GP model is to identify a best choice of
the distribution's ${\bf x}$-dependent mean $\mu^{(n)}(\mathbf x)$ and width
$\sigma^{(n)}(\mathbf x)$ given information available at iteration $n$.  In
practice this problem is still intractable, so to reduce the search space GP
models employ correlation length hyperparameters that describe the range in
every direction of parameter space over which functions are likely to exhibit
correlations.  At each iteration, a GP model generally uses the most likely
function $C^{(n)}_{\rm GP}({\bf x}) = \mu^{(n)}({\bf x})$ as the estimated cost
function.

Given a GP model at iteration $n$, a numerical optimization algorithm
determines the next parameter vector $\mathbf{x}^{(n+1)}$ to be sampled.  One
common choice is the upper confidence bound criterion: $\mathbf x^{(n+1)} =
\mathrm{argmin} \, \{\mu ({\mathbf x}^{(n)}) + \alpha \sigma({\mathbf
x}^{(n)})\}$, where the hyperparmeter $\alpha$ balances pursuing the minimum
and exploring the unknown region of the model and is dynamically regulated
during the optimization procedure.  The potential optimal $\mathbf{x}^{(n+1)}$
is experimentally measured, yielding $C_m(\mathbf{x}^{(n+1)})$, the GP
parameters are updated, and the process repeats with a new parameter vector.
This iterative procedure continues until a termination criterion is met, such
as a predefined number of iterations or convergence to an optimal parameter
set.

We employ the open-source package `Machine-Learning Online Optimization
Package' ({\tt M-LOOP}) \cite{wigley2016fast} to implement the GP optimization
and interface it with our {\tt labscript}-based\cite{Starkey2013} experiment
control system.  {\tt M-LOOP} is seeded with an initial dataset obtained by
experimentally sampling the cost at parameter vectors $\mathbf{x}^{(n)}$.  The
set of vectors might be selected: at random, with a different optimizer like
differential evolution (our choice), or even with a previous GP model.  The
measured points help to update GP model's hyperparameters.  The GP optimizer
allows us to efficiently explore the multiparameter space by prioritizing
regions with higher uncertainty, unlike regular grid scans that uniformly
sample the space.  This allows faster convergence to optimal parameters
compared to the conventional scan-based approach in fewer runs.

We note that the samples collected during GP optimization are insufficient to
accurately extract the correlation matrix in the region surrounding the cost
extrema; this is due to the low sample count required for convergence.

\subsection{Constrained optimization} \label{sec:constrained_optimization}

\begin{figure}[t]

\includegraphics{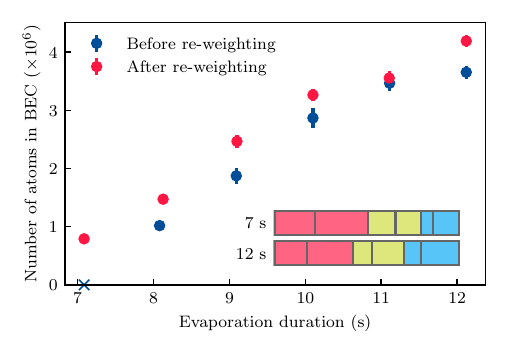}

\caption{
    Evaporation sequence performance.  The blue/red dots mark the performance
    of the initial sequence before/after re-weighting.  The blue cross at 7~s
    indicates that the sequence before re-weighting utterly failed in producing
    a BEC.  The two bars in the inset show the percentage of time spent in each
    stage for the optimal 7~s and 12~s sequence.  The red, green, and blue
    region marks the rf evaporation stage, the decompression/transfer stage,
    and the hybrid trap evaporation stage, respectively.
}

\label{fig:evaporation_comparison}

\end{figure}

Our constrained optimization method is built upon the standard Lagrange
multiplier (LM) technique.  While this approach is analytically
straightforward, its implementation in a data-sparse setting with experimental
noise requires special consideration.

For the case of a single constraint $f(\mathbf x) = 0$ the LM method first
introduces a new cost function $C_{\text{LM}}(\mathbf x) = C(\mathbf x) +
\lambda f(\mathbf x)$, where the coefficient $\lambda$ is the LM.  As usual,
local minima of $C_{\text{LM}}$ occur at points of zero gradient \begin{align}
    \nabla_{\mathbf x} C_{\text{LM}}(\mathbf x) &= \nabla_{\mathbf x} C(\mathbf
    x) + \lambda \nabla_{\mathbf x} f(\mathbf x) = 0. \label{eq:langrange_cond}
\end{align} In conjunction with the constraint $f(\mathbf x) = 0$, this yields
$n+1$ coupled equations: sufficient to resolve the $n={\rm dim}({\bf x})$
components of ${\bf x}$ along with the LM $\lambda$.  The gradient of the
constraint $\nabla_{\mathbf x} f(\mathbf x)$ can be directly evaluated
analytically since the constraint is known.  In contrast, $\nabla_{\mathbf x}
C(\mathbf x)$ is an experimental quantity which can be estimated either by line
scans or more efficiently via random sampling in the neighborhood of ${\bf x}$.

The strategy for updating the parameter vector $\mathbf x^{(n)}$ at iteration
$n$ to $\mathbf x^{(n+1)}$  proceeds as follows.  First, we estimate the cost
function and its gradient at $\mathbf x^{(n)}$ using a sampling strategy as
discussed in Sec.~\ref{sec:parameter_grouping} above, and obtain the gradient
$\nabla_{\mathbf x} f(\mathbf x^{(n)})$ from a linear fit to the data.  We then
employ a constrained gradient descent to update $\mathbf x^{(n+1)} = \mathbf
x^{(n)} + \beta {\bf P}_f \nabla_{\mathbf x} C(\mathbf x^{(n)})$ with
convergence parameter $\beta$ and ${\bf P}_f$ projects onto the surface where
$f(\mathbf x) = 0$.  Separately, we solve the $n$ linear equations in
Eq.~\eqref{eq:langrange_cond}, each with a separate $\lambda_i$ as the unknown
quantity.  These $\lambda_i$'s will be equal only when $\mathbf x^{(n)}$ is an
exceptional point of the constrained optimization problem.  The optimization
therefore terminates when $|\lambda_i - \lambda_j|< \epsilon$ for all $i,j$.

We demonstrate this procedure by optimizing the duration of each evaporation
stage while constraining their sum: the total evaporation time $T$.  Without
the constraint, the optimal evaporation sequence would balance improved
thermalization (favoring longer durations) and vacuum-limited lifetime
(favoring shorter durations).  Constraining the total duration therefore
strategically exchanges time amongst stages in a way that optimizes BEC
production.

For our specific linear constraint $f(t_1, \dots, t_n) = \sum_i t_i - T = 0$,
Eq.~\eqref{eq:langrange_cond} gives $\lambda_i = - \partial_{t_i} C(t_1, \dots,
t_n)$.  The gradient descent process can be thus be reframed as a re-weighting
process in which the duration of stages with small $\lambda_i$ are reduced,
while the duration of stages with large $\lambda_i$ are increased.  This
process is rapidly convergent, usually within three iterations; we tune $\beta$
at each step to ensure $\sum_{j} |t_j^{(n)} - t_j^{(n+1)}| = 0.6 \, \mathrm s$
and terminate when $|\lambda_i - \lambda_j|$ is smaller than the largest
uncertainties in all $\lambda_i$'s.  In practice, we perform a final stage of
GP optimization (with durations) after completing the constrained optimization
process, however, at best this yields a marginal improvement.

Figure~\ref{fig:evaporation_comparison} shows performance improvement of this
process (red) compared to simply proportional scaling the stage durations
(blue) for sequences with constrained evaporation durations from $7\ {\rm s}$
to $12\ {\rm s}$.  The bar-graphs in Fig.~\ref{fig:evaporation_comparison} show
how this process fractionally redistributes time between the different stages.
As the evaporation duration is reduced the fraction of time allocated to hybrid
trap evaporation (blue) is reduced and redistributed to the rf evaporation
stage (red).

We fully automated this process using our {\tt labscript}-based experiment
control system, with supporting code available online \footnote{Code is
available at
\url{https://github.com/JQIamo/labscript-time-reweight-optimization}}.

\section{Conclusion}\label{sec:conclusion}

Here we described an apparatus that produces large sodium BECs in a hybrid
trap, and describe efficiency gains achieved via parameter grouping,  GP
optimization, and LM-based constrained optimization.  This integrated approach
strikes a practical balance between performance of the optimization system
while still benefiting from correlations between parameters.  This enabled our
BEC apparatus to outperform existing sodium BEC systems in terms of BEC
production efficiency as a function of evaporation time (see Fig.
\ref{fig:apparatus_compare}).  We also note that improvements in laser cooling,
such as D1 gray molasses~\cite{PhysRevA.93.023421}, could boost the phase space
density prior to evaporation and circumvent the need for large-volume magnetic
traps (and their attendant slow evaporation) for the production of large BECs.

Our automated time re-weighting protocol, based on constrained optimization,
efficiently redistributes the evaporation sequence under fixed time
constraints, significantly enhancing BEC quality by shifting resources from
less effective to more impactful stages.  In the results presented here, we
minimized the sequence duration with fixed MOT parameters, arguing that the MOT
and evaporation stages are largely decoupled.  In reality these stages are
correlated (albeit weakly); this correlation can be leveraged to even more
effectively reduce the total sequence duration.  For example, increasing the
initial number of laser cooled atoms increases the collision rate in the
magnetic trap, allowing for more rapid evaporation.  So in such a situation,
our LM technique may well sacrifice evaporation time for added MOT loading
time.

\begin{acknowledgments}

The authors thank K.~T.~Hoang and R.~Shrestha for carefully reading the
manuscript, and J.~V.~Porto, S. Subhankar for initial discussions.  This work
was partially supported by the National Institute of Standards and Technology,
and the National Science Foundation through the Quantum Leap Challenge
Institute for Robust Quantum Simulation (grant OMA-2120757).
\end{acknowledgments}

\appendix \section{All machine-optimized parameters}

All parameters optimized by the Bayesian optimizer are listed in
Table~\ref{tab:params}.

\begin{table*}[t] \centering \begin{tabular}{lcl} \firsthline Experiment stage
& Parameter & Description                                \\ \hline 3D MOT
loading                                            & $I_{\rm MOT}$ & $F=2
\rightarrow F'=3$ cooling intensity                  \\
							   & $\delta_{\rm MOT}$
							   & $F=2 \rightarrow
							   F'=3$ cooling
							   detuning
							   \\ & $B_{x,y,z}$ &
							   Bias fields
							   \\ & $B'$ &
							   Quadrupole magnetic
							   field gradient
							   \\
\hline PGC                                                        & $I_{\rm
PGC}$ & $F=2 \rightarrow F'=3$ cooling intensity                  \\
							   & $\delta_{\rm PGC}$
							   & $F=2 \rightarrow
							   F'=3$ cooling
							   detuning
							   \\ & $B_{x,y,z}$ &
							   Bias fields
							   \\ &$ t_{\rm PGC}$ &
							   PGC duration
							   \\
\hline Optical pumping                                            & $I_{\rm
depump}$ & $F=2 \rightarrow F'=2$ depump intensity                  \\
							   & $\delta_{\rm
							   depump}$ & $F=2
							   \rightarrow F'=2$
							   depump detuning
							   \\ & $I_{\rm pump}$
							   & $F=1 \rightarrow
							   F'=1$ optical
							   pumping intensity
							   \\ & $B_{x,y,z}$ &
							   Bias fields
							   \\
\hline Magnetic trap compress                                     & $B'_{\rm
capture}$ & Capture quadrupole magnetic field gradient \\
							   & $B'_{\rm 0}$ &
							   Final compressed
							   quadrupole magnetic
							   field gradient \\ &
							   $t_{\rm compress}$ &
							   Compression duration
							   \\ & $f_{{\rm
							   rf},0}$ & Initial RF
							   frequency
							   \\ & $I_{{\rm
							   ODT},0}$ & Initial
							   ODT power
							   \\
\hline Evaporation stage $j=1,\cdots,6$.                          & $f_{{\rm
rf}, j}$ & Ending RF frequency                                         \\
							   & $B_{[x,y,z],j}$ &
							   Ending Bias fields
							   \\ & $I_{{\rm
							   ODT},j}$ & Ending
							   ODT power
							   \\ &
							   $B^{\prime}_{j}$ &
							   Ending Magnetic
							   quadrupole field
							   gradient
							   \\ & $t_{j}$ & Time
							   duration
							   \\
\lasthline \end{tabular} \caption{List of 53 parameters that were
machine-optimized as part of this study.} \label{tab:params} \end{table*}

\bibliography{main}

\end{document}